\documentclass[11pt,aps,prd,amsmath,amssymb,superscriptaddress,nofootinbib]{revtex4-1}
\pdfoutput=1
\usepackage{amsmath,amssymb} 
\usepackage{epsfig} 
\usepackage{color}
\usepackage{bm} 
\usepackage{rotating}
\usepackage{physics}
\usepackage{braket}
\usepackage{multirow}
\usepackage{hhline}
\usepackage{colortbl}
\usepackage[table]{xcolor}
\usepackage{booktabs}
\usepackage{graphicx}
\usepackage[left=2.5cm,right=2.5cm,top=2cm,bottom=2cm]{geometry}
\newcommand{\X}{X(3872)}

\begin{document}

\title{Radiative decay of the $X(3872)$ as a mixed molecule-charmonium state in effective field theory }

\author{E. Cincioglu}
\email{elifc@metu.edu.tr}
\affiliation{Department of Physics, Middle East Technical University, Ankara, Turkey}
\author{A. Ozpineci}
%\email{ozpineci@metu.edu.tr}
\affiliation{Department of Physics, Middle East Technical University, Ankara, Turkey}

\date{\today}

\begin{abstract}
Assuming that $X(3872)$ is a mixture between $2P$ charmonium and $\bar{D}D^{*}$ molecular states with $J^{PC}=1^{++}$, an analysis of $X(3872)$ radiative decays into $J/\psi \gamma$ and $\psi(2S)\gamma$ is presented. 
The modification of  the radiative branching ratio due to possible constructive or destructive interferences between the meson-loop and the short-distance contact term, which is modeled by a charm quark loop, is shown.  The model predictions are shown to be compatible with the experimentally determined ratio of the mentioned branching fractions for a wide range of the $X(3872)$ charmonium content. In the case of the destructive interference, a strong restriction on the charmonium admixture is found.
\end{abstract}

\maketitle

\section{Introduction}
The $X(3872)$ state was first observed by Belle \cite{Choi:2003ue} through the channel $B^{\pm}\to J/\psi\pi^{+}\pi^{-}K^{\pm}$ and its quantum numbers were determined as $1^{++}$ \cite{Aaij:2013zoa}.  The averaged mass of the $X(3872)$ is $3871.69\pm 0.17$ MeV, and the full width is small, $\Gamma <1.2$ MeV, which is not easily accommodated in the potential quark models. Moreover, its mass does not fit into the traditional quark model as non-relativistic bound state of charm quarks. Despite the other possibilities including a molecular state consisting of a $D$ and $\bar{D}^{*}$ \cite{Gamermann:2007fi, Liu:2008fh, Liu:2007bf, Dong:2008gb, Swanson:2003tb, Voloshin:2004mh, Braaten:2005ai, Gamermann:2009uq}, tetraquark \cite{Maiani:2004vq, Dubnicka:2010kz, Dubnicka:2011mm}, $c\bar{c}-D\bar{D}^{*}$ mixing \cite{Badalian:2012jz, Wang:2010ej, Eichten:2005ga, Dong:2009uf} or radial excitation of the $P-$wave charmonium \cite{Barnes:2005pb}, the structure of the $X(3872)$ is not yet fully understood. Since the mass of the $X(3872)$ is extremely close to the $D^0\bar{D}^{*0}$ threshold, many authors have suggested that it is a loosely bound state of $D\bar{D}^{*}$. In addition, predominantly molecular description of $X(3872)$ is also favored by the experimental ratio \cite{Abe:2005ix} of decay fractions of $X(3872)$ into $J/\psi \pi^{+}\pi^{-}$ and $J/\psi \pi^{+}\pi^{-}\pi^{0}$ final states \cite{Gamermann:2009fv,Gamermann:2009uq}. 

Another puzzling observation about $X(3872)$ is its radiative decays. The ratio of the branching fractions into  final states with a photon and a $J/\psi$ or $\psi(2S)$ has been measured \cite{Aubert:2008ae, Aaij:2014ala} as 
\begin{equation}
\label{eq:Rexprad}
R_{\psi\gamma}=\frac{B_{r}(X\to\psi(2S)\gamma)}{B_{r}(X\to J/ \psi \gamma)}=2.46\pm 0.64\pm 0.29.
\end{equation}
Various quark model calculations describing the $X(3872)$ as a radially excited $\chi_{c1}(2P)$ charmonium state predict a wide range of values for this ratio. However, the results are very sensitive to quark model details since the radiative decay matrix element is proportional to the overlap integral of the initial state and the final state wave functions.  An alternative discussion is presented in the work of Swanson et al.~\cite{Swanson:2004pp}, where using vector meson dominance, it is argued that if $X(3872)$ is a predominantly molecular state, the ratio is predicted as $4\times10^{-3}$ which is three orders smaller than the observed ratio. Contrary to the claim in this study, in Ref~\cite{Guo:2014taa} it was demonstrated that the observed ratio allows the $X(3872)$ to be a hadronic molecule with the dominant component $D\bar{D}^{*}$. In addition, the production rate of $X(3872)$ in the $p\bar{p}$ collisions which is about $1/20$ of the rate of $\psi{(2S)}$ can easily  be accommodated with an admixture of approximately $5\%$ of a $c\bar{c}$ component in the its wave function. The charmonium admixture in a molecular picture of $X(3872)$ has been studied in Ref.~\cite{Dong:2009uf}. There it was concluded that the observed ratio can be explained, if one assumes that the compact component of the $X(3872)$ is $5-12\%$.

Within the molecular description of the $X(3872)$, triangular $DD^{(*)}\bar{D}^{(*)}$ and simple $D\bar{D}^{*}$ loop contributions to the radiative amplitude, without explicitly considering the short-range contributions, were computed in Ref.~\cite{Guo:2014taa}. In exploratory study of Ref.~\cite{Cincioglu:2016fkm}, the size of the counter-term was estimated in an effective field theory framework allowing for both a molecular as well as a compact component of the $X(3872)$. However, the meson loop contribution in the $X(3872)\to J/\psi\gamma$ mode and possible interferences effects between the meson-loop and the counter-term contributions in both of the decays are neglected. Moreover, it was claimed in Ref.\cite{Dong:2009uf} that the relative phases of the coupling constants are uncertain and they can be fixed by an analysis of the branching ratio data. 

In this study, we investigate the effects of short-range contributions to the radiative decays of the $X(3872)$ into $\psi(2S)\gamma$ and $J/\psi\gamma$ in an effective field theory allowing a $\chi_{c1}(2P)$ charmonium admixture in the molecular state. We demonstrate that the relative phase of the couplings are important to determine whether the charmonium content of the $X(3872)$ is nontrivial.

\section{Formalism}
As mentioned in the introduction, the triangular $DD^{(*)}\bar{D}^{(*)}$ and simple $D\bar{D}^{*}$ loop contributions to the radiative decays were calculated from diagrams Fig.1(a-e) in the work by Guo et al.~\cite{Guo:2014taa} within an effective theory framework.  In \cite{Guo:2014taa} , the contributions to the loop amplitude from these diagrams are written as 
\begin{equation}
\mathcal{M}^{loop}_{\mu\sigma\lambda}=\frac{1}{\sqrt{2}}e g_{XDD^{*}}g_{\psi DD^{*}} m \sqrt{m_Xm_{\psi}}\int{\frac{d^{4}}k}{4\pi}S^{\nu\sigma}(k)S(k-p)J_{\mu\nu\lambda}(k),
\label{amp.loop}
\end{equation}
where $J_{\mu\nu\lambda}$ tensor includes the electric and magnetic contributions and $S^{\nu\sigma}(k)$ and $S(k-p)$ are the $D^{*}$, $D$ propagators, respectively.
The couplings 
 $g_2$ and $g_{2}^{\prime}$ are used  for the spin symmetric couplings of  $\psi(nS)D^{(*)}\bar{D}^{(*)}$, $g_{\psi DD^{*}}$, for $n=1$ and $n=2$ respectively. Finally, the coupling constant of the $X(3872)$ to $\bar{D}D^{*}$, $g_{XDD^{*}}$, can be expressed as follows in terms of the probability, $\tilde Z_{\X}$,  to find the molecular component $\bar{D}D^{*}$ in the physical wave function of the $X(3872)$ \cite{Cincioglu:2016fkm}
\begin{equation}
g_{XDD^{*}}=\left(\frac{-f_{\Lambda}^{2}}{G_{QM}^{\prime}} \tilde Z_{\X} \right)^{1/2},
\label{gDDX}
\end{equation}
where $f_{\Lambda}$ is Gaussian regulator for on-shell mesons which depends on the masses of the involved mesons and $G_{QM}^{\prime}$ is derivative of the meson-loop function with respect to energy. For the numerical analysis, the coupling constant, $g_{XDD^*}$ are taken  from Table I of Ref.~\cite{Cincioglu:2016fkm}.
\begin{figure}[h!]
\begin{center}
\includegraphics[width=0.8\textwidth]{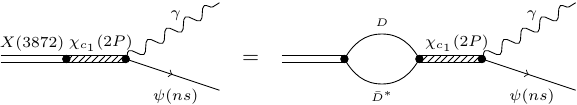}
\caption{Decay mechanism for the transition $\X\to \psi(nS)$ through
  an intermediate charmonium $\chi_{c1}(2P)$ state \cite{Cincioglu:2016fkm} }
 \label{fig:counterterm}
\end{center}
\end{figure}

Since the loop integral in the amplitude (Eq.\ref{amp.loop}) is divergent, one needs to include a counter-term to renormalize the ultraviolet divergences of the loop diagrams. After the renormalization procedure, the counter-term modeled by a charm quark loop in Fig.~\ref{fig:counterterm} provides a finite contribution to the total decay width. To estimate the strength of the short range interaction we use the effective field theory approach of Ref.~\cite{Cincioglu:2016fkm}  which incorporates possible mixing between the molecular $D\bar{D}^{*}$ and $\chi_{c1}(2P)$ charmonium state. The contribution from short range interaction depicted in diagram Fig.~\ref{fig:counterterm} can be obtained as 
\begin{equation}
\mathcal{A}_{\mu\sigma\lambda} =-i\sqrt{2}(\tilde Z_{\X}\times f(\tilde Z_{\X})^{1/2}m_{X} m_{\psi}\delta^{nS2P}v_{\eta}\epsilon^{\sigma\mu\rho\eta}(v.qg^{\rho\lambda}-q^{\rho}v^{\lambda}),
\label{amp.charmonium}
\end{equation}
where $ f(\tilde Z_{\X})$ is the dressed and bare charmonium propagator ratio squared \cite{Cincioglu:2016fkm}. As in seen in Eq.~\ref{amp.charmonium}, $\mathcal{A}_{\mu\sigma\lambda}$ depends on the $\chi_{c1}(2P)\psi(nS)\gamma$ coupling  $\delta^{nS2P}$. This coupling is one of the greatest uncertainties of the present calculation. It can be written as
\begin{equation}
\delta^{nS2P}= \left(\frac{4\pi\alpha e_{c}^{2}}{3}\right)^{1/2}\mel{\psi(nS)}{r}{\chi_{c1}(2P)},
\end{equation}
where $e_c$ is the charm quark electric charge, $\alpha$ is the fine-structure constant and the overlap integral of the initial state and the final state wave functions $\mel{\psi(nS)}{r}{\chi_{c1}(2P)}$ can be calculated by using quark model wave functions.

Finally, the total amplitude can be expressed as follows
\begin{equation}
\mathcal{M}^{full}=(\mathcal{A}_{\mu\sigma\lambda}+\mathcal{M}^{loop}_{\mu\sigma\lambda})
\epsilon_{(X)}^{\sigma}(p)\epsilon_{(\psi)}^{\mu}(p-q)\epsilon_{(\gamma)}^{\lambda}(q).
\label{amp.full}
\end{equation}

\subsection{General remarks}

As pointed out in Ref.~\cite{Guo:2014taa}, quark model calculations predict a wide range for the radiative branching fractions $R_{\psi \gamma}$ assuming a $\chi_{c1}(2P)$ $c\bar{c}$ nature for the $X(3872)$, where the decay width results are very sensitive to quark model details in particular in the $J/\psi$ mode. The $\delta^{nS2P}$ couplings used in Ref.~\cite{Cincioglu:2016fkm} are based on non-relativistic quark model of Ref.\cite{Barnes:2005pb}. Here, two different quark model estimates for the  overlap integrals, Set 1 and Set 2, of the initial state and the final state wave functions are considered to see dependence of the predictions presented in this work on the coupling constants, $\delta^{nS2P}$, (see Table~\ref{tab:Rad.decays})
%%%%
\begin{table}[h!]
\centering
\begin{tabular}{l|ccccc}
& Final& $ \mel{\psi(nS)}{r}{\chi_{c1}(2P)}$ & $\delta^{nS2P}$ & $\Gamma[X(3872)\to\psi(nS)\gamma]$ &  $R_{\psi\gamma}$   \\
& state&$[{\rm GeV}^{-1}]$ &$[{\rm GeV}^{-1}]$ &$[{\rm keV}]$ &   \\ \hline \hline
\multirow{2}{*}{Set 1} &$J\psi \gamma$ &$0.389^{\dagger}$&0.045& 60.87& \multirow{2}{*}{1.48} \\
& $\psi(2S)\gamma$ &$3.26^{\dagger}$& 0.38& 87.25& \\ \hline 
\multirow{2}{*}{Set 2}&$J\psi \gamma$ &0.202~\cite{Takizawa:2012hy}&0.024& 16.06& \multirow{2}{*}{3.52} \\
& $\psi(2S)\gamma$ &2.63\cite{Takizawa:2012hy}& 0.31& 56.65& \\ \hline \hline
\end{tabular}
\caption{ Radiative decays of $X(3872)$ as a radially excited $\chi_{c1}(2P)$ charmonium state are based on quark model estimates. $^\dagger$:Overlap integrals are estimated from the widths given in Table III of Ref.~\cite{Barnes:2005pb} for the $\chi_{c1}(2p)\to\psi(nS)\gamma$ $E_1$ radiative transitions calculated by the non-relativistic quark model.} 
\label{tab:Rad.decays}
\end{table}

In the analysis of Ref.~\cite{Guo:2014taa}, the dependence of the radiative branching ratios on the  coupling of $X(3872)$ to the charmed mesons, $g_{XDD^{*}}$, cancels in the ratio, since only the loop contributions are considered. However, in the present work, since the charm quark loop contribution does not contain this coupling, the predicted ratio depends on it. For the numerical analysis, the values of the $g_{XDD^{*}}$ are taken from Table I of Ref.~\cite{Cincioglu:2016fkm}. Moreover, the ratio $R_{\psi\gamma}$ obtained in  Ref.~\cite{Guo:2014taa} depends on the ratio of the couplings $r_{g_2,g_{2}^{\prime}}$, while it is separately dependent on  $g_2$ and $g_{2}^{\prime}$ in the study presented here. In Ref.~\cite{Dong:2009uf},  it was found that $r_{g_2,g_{2}^{\prime}}\simeq2$. In addition, using vector dominance arguments, the coupling constant, $g_2$, of $J/\psi$ to the charm meson-antimeson pair was estimated as about 2 GeV$^{-3/2}$ in Ref.~\cite{Guo:2010ak}. Model independent estimates for that coupling were given in a range of $2.1-2.9$ GeV$^{-3/2}$ in Ref.~\cite{Matinyan:1998cb, Deandrea:2003pv, Matheus:2002nq}. In line with these considerations, in the present work, $g_2$ is taken as 2.5  GeV$^{-3/2}$, and the results are analyzed for various values of $r_{g_2,g'_2}$. 

On the other hand, the loop integrals in Eq.~\ref{amp.loop} are scale dependent. In Eq.~\ref{amp.full}, the cut-off dependence of the $\mathcal{M}^{loop}_{\mu\sigma\lambda}$ should be compensated by a corresponding variation in the counter-term contribution $\mathcal{A}_{\mu\sigma\lambda}$. Here, we have computed the full amplitude $\mathcal{M}^{full}$ using dimension regularization with the $\overline{MS}$ subtraction scheme, while the couplings of $X(3872)$ state to the charmonium and $D\bar{D}^{*}$ molecule were computed in Ref.~\cite{Cincioglu:2016fkm} using an ultraviolet cut-off at the scale $\Lambda=1$ GeV. In Ref.~\cite{Cincioglu:2016fkm}, both of the regularization schemes were compared considering the two meson-loop function and found that UV cut-off at the scale $\Lambda=1$ GeV would correspond to a $\overline{MS}$ scale, $\mu$ of the order of 1 GeV. Therefore, all calculations have been carried out with $\overline{MS}$ scale, $\mu=1$ GeV. 

Finally, the importance of the relative signs of the coupling constants was stressed in Ref.\cite{Dong:2009uf}. To study the effects of this phase, along the lines of study in Ref.\cite{Dong:2009uf},   the coupling $g_2\,(g_{2}^{\prime})$ is given an arbitrary phase $e^{i\phi}$ with $0 \leq \phi \leq \pi$. 

\section{Results and Discussion}

\begin{figure}[h!]
\centering
\begin{tabular}{ccc}
\includegraphics[width=0.5\textwidth]{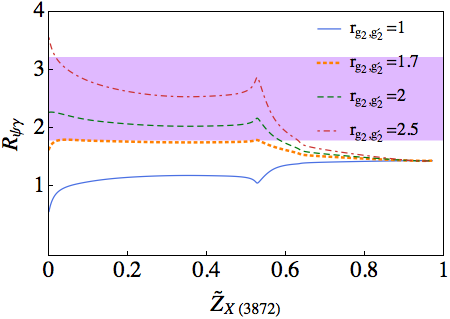} &
\includegraphics[width=0.5\textwidth]{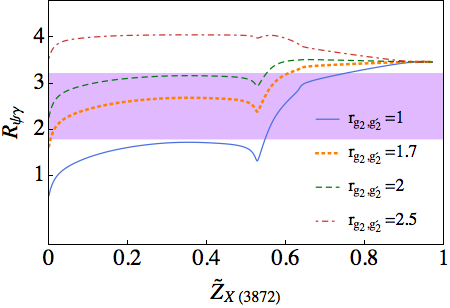} \\
\includegraphics[width=0.5\textwidth]{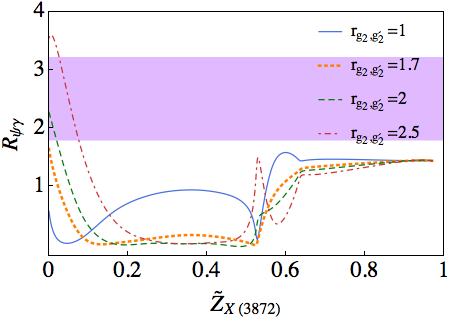} &
\includegraphics[width=0.5\textwidth]{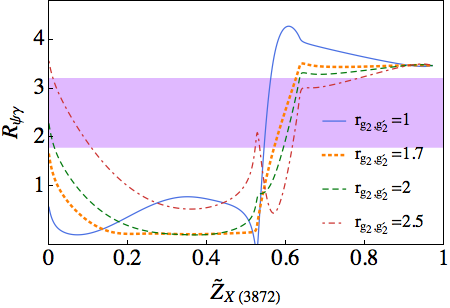} \\
\includegraphics[width=0.5\textwidth]{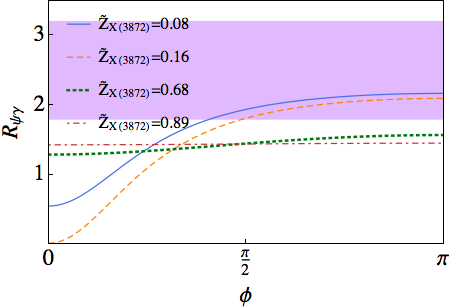} &
\includegraphics[width=0.5\textwidth]{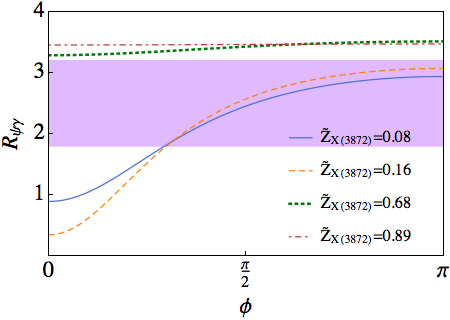}
\end{tabular}
\caption{$R_{\psi\gamma}$ as a function of the charmonium probability $\tilde Z_{\X}$ (first two rows) and $\phi$ (last row). See text for explanation of the legends.}\label{fig:radZtilde}
\end{figure}

In Fig.~\ref{fig:radZtilde}, the ratio $R_{\psi\gamma}$ is shown as a function of the $X(3872)$ charmonium content, $\tilde Z_{\X}$, and $\phi$.
In the first two rows, the dependence of $R_{\psi\gamma}$ is depicted for four different values of the coupling constant ratio $r_{g_2,g_{2}^{\prime}}=1$, $1.7$, $2$ and $2.5$, along with the experimental band. In the first row $\phi$ is set to $\phi=\pi$ (constructive interference) and in the second row $\phi=0$ (destructive interference). The third row depicts the dependence of $R_{\psi\gamma}$ on the phase $\phi$ for $\tilde Z_\X=0.08$, $0.16$, $0.68$ and $0.89$ with $r_{g_2,g'_2}=2$. In the first column, the values of $\delta^{nS2P}$ are taken from Set 1, and in the second column, they are taken from Set 2. Note that, as can be seen from Table~\ref{tab:Rad.decays}, $\delta^{2S2P}$ is almost the same in the two quark models. Hence, the difference in the figures in the two rows are mainly due to the different values of $\delta^{1S2P}$.

As can be seen from the figure, contrary to the findings of Ref.~\cite{Dong:2009uf}, both a trivial and a non-trivial charmonium component of $\X$ is consistent with the radiative decay ratio, independent of which quark model prediction is used for the coupling constants $\delta^{nS2P}$. 

It can be also seen from the figure that the behavior of the predictions of the ratio of radiative decays  is different when $\tilde Z_\X \lesssim 0.55$ and when $\tilde Z_X \gtrsim 0.55$. For large values of $\tilde Z_\X$, the prediction of the ratio has a small dependence on the ratio $r_{g_2,g'_2}$ and $\phi$, and a large dependence on $\delta^{nS2P}$. This is the expected behavior, since in this range, $\X$ is dominantly a charmonium state. Such large values of $\tilde Z_\X$ can be consistent with the observed ratio, provided that $\delta^{nS2P}$ has a value in between  the values of $\delta^{nS2P}$ used in this work. For smaller values of $\tilde Z_\X \lesssim 0.55$, the predictions are less sensitive to $\delta^{nS2P}$, but more sensitive to $r_{g_2,g'_2}$ and $\phi$\footnote{Note that, when $\tilde Z_\X=0$, there is no $\phi$ dependence. Hence for very small values of $\tilde Z_\X$, the dependence on $\phi$ is also small.}. In the case of constructive interference, the first row of figures, the predicted ratio is consistent with experimental values for both values of $\delta^{nS2P}$ for almost any value of $\tilde Z_\X \lesssim 0.55$ if $r_{g_2,g'_2}$ is between $1.7$ and $2$. Larger values of $r_{g_2,g'_2}$ would become consistent with observations for larger values of  $\delta^{nS2P}$, and smaller values of $r_{g_2,g'_2}$ are consistent with observations for smaller values of $\delta^{nS2P}$. In the case of destructive interference, the second row of figures, the allowed range of $\tilde Z_\X$ is pushed to  smaller values. 

From the above considerations, it is concluded that a wide range of charmonium probability in the $X(3872)$ is consistent with the experimentally observed value of $R_{\psi\gamma}$. This confirms that this ratio is not in conflict with a predominantly molecular or charmonium nature of the $X(3872)$. In the case of the destructive interferences between the meson loops and the counter-term, a strong constraint on the $\chi_{c1}(2P)$ content in the $X(3872)$ is found. Actually, it is expected that charmonium contents of the $X(3872)$ is smaller than $15\%$ because a significantly larger $c\bar{c}$ content may not explain the experimental ratio of decay fractions of $X(3872)$ into $J/\psi\pi^{+}\pi^{-}$ and $J/\psi\pi^{+}\pi^{-}\pi^{0}$ final states~\cite{Cincioglu:2016fkm}. A detailed analysis of isospin violation in the decays of $X(3872)$ and a more precise knowledge of $\delta^{2S2P}$ coupling would put a stronger restriction on the charmonium admixture in the $X(3872)$.

\subsection*{Acknowledgement}
This research has been supported by TUBITAK (The Scientific and Technological Research Council of Turkey) under the grant no 114F234.
\newpage
\bibliography{paper}

\end{document}